\documentclass[11pt]{article}
\usepackage[margin=1in]{geometry}
\usepackage{graphicx}
\usepackage{natbib}
\usepackage{enumerate,amsmath,amssymb,amsfonts,amsthm,subcaption,hyperref}
\usepackage[linesnumbered,ruled]{algorithm2e}
\newcommand{\N}{{\mathbb N}}
\newcommand{\R}{{\mathbb R}}

\begin{document}
\title{Peering into the Anneal Process of a Quantum Annealer}
\author{Elijah Pelofske \and Georg Hahn \and Hristo Djidjev}
\date{Los Alamos National Laboratory}
\maketitle

\begin{abstract}
Commercial adiabatic quantum annealers have the potential to solve important NP-hard optimization problems efficiently. The newest generation of those machines additionally allows the user to customize the anneal schedule, that is, the schedule with which the anneal fraction is changed from the start to the end of the annealing. In this work we use the aforementioned feature of the D-Wave 2000Q to attempt to monitor how the anneal solution evolves during the anneal process. This process we call slicing: at each time slice during the anneal, we are able to obtain an approximate distribution of anneal solutions. We use our technique to obtain a variety of insights into the D-Wave 2000Q. For example, we observe when individual bits flip during the anneal process and when they stabilize, which allows us to determine the freeze-out point for each qubit individually. We highlight our results using both random QUBO (quadratic unconstrained binary optimization) instances and, for better visualization, instances which we specifically optimize (using our own genetic algorithm) to exhibit a pronounced evolution of its solution during the anneal.
\end{abstract}

\section{Introduction}
\label{sec:intro}
Quantum computers of D-Wave Systems, Inc., make it possible to obtain fast approximate solutions of NP-hard problems of very high quality using a process called \textit{quantum annealing} \cite{D-WaveSystems2000QuantumToday}. 
The D-Wave annealer is designed to minimize a quadratic  function $Q$ given by
\begin{align}
    Q({q_1,\ldots,q_n} )=\sum_{i=1}^n a_i q_i + \sum_{i\leq j} a_{ij} q_i q_j,
    \label{eq:hamiltonian}
\end{align}
where $n \in \N$ is the number of variables, $a_i \in \R$ are the linear weights, and $a_{ij} \in \R$ are the quadratic weights. Depending on the values the variables can take, the function in \eqref{eq:hamiltonian} is called an \textit{Ising}  if $q_i \in \{-1,+1\}$, or a \textit{QUBO} (quadratic unconstrained binary optimization) problem if $q_i \in \{0,1\}$. Both formulations are equivalent \cite{Chapuis2019}. In this  article, we solely focus on the QUBO formulation. Many important NP-complete problems, such as the maximum clique, minimum vertex cover,  or the graph coloring problem, can be expressed as a minimization of the form \eqref{eq:hamiltonian}, see \cite{PelofskeQTOP19,PelofskeCompFront19,DwaveMapColoring,Lucas2014}.

Typically, in order to use D-Wave to obtain a high quality solution of an NP-complete problem, we proceed in three steps: First, the problem under investigation, for example an instance of the Maximum Clique problem, is expressed as a minimization of type \eqref{eq:hamiltonian}. Second, the coefficients $a_i$ and $a_{ij}$ of an instance of \eqref{eq:hamiltonian} are mapped onto the qubits and the connections between them (called couplers) of the D-Wave chip \cite{TechnicalDescriptionDwave}. Once this is done, the third step is to request a pre-specified number of anneals (solutions) from the D-Wave annealer. Importantly, the sole output of an anneal is a bit string indicating the value of each variable $q_i$ read off the $i$-th qubit at the end of the anneal process.

\begin{figure}
    \centering
    \includegraphics[width=0.5\textwidth]{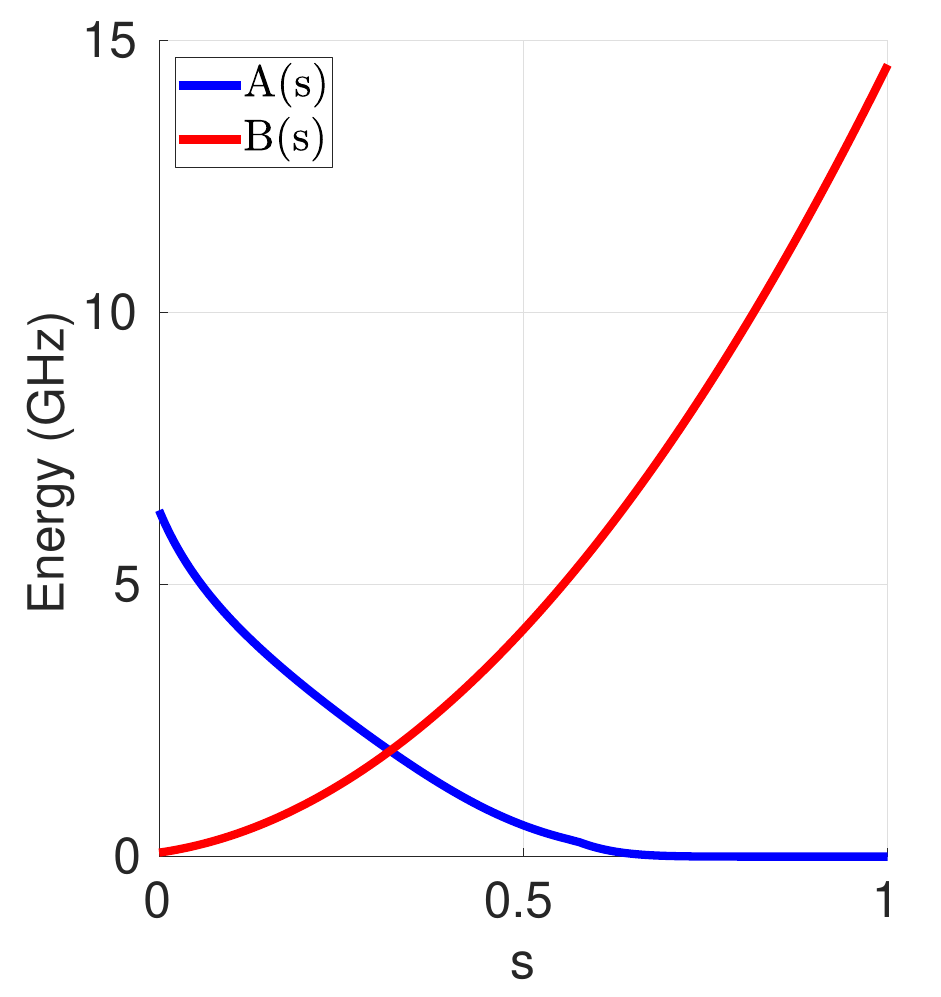}
    \caption{Functions $A(s)$ and $B(s)$ for D-Wave 2000Q \cite{LANLDwave}, where $s \in [0,1]$ is the annealing fraction.}
    \label{fig:anneal_schedule}
\end{figure}

The time evolution of any quantum system can be described by an operator called \textit{Hamiltonian}, and the D-Wave processor evolves a quantum system with a time-dependent Hamiltonian 
$$H(s)=-\frac{A(s)}{2}\sum_{i=1}^n a_i\sigma^x_i +\frac{B(s)}{2}\Big(\sum_{i=1}^n\sigma^z_i + \sum_{i\leq j} a_{ij} \sigma^z_i \sigma^z_j\Big),$$
whose first term represents a quantum state where each output bit string is equally likely, and whose second term encodes the coefficient from \eqref{eq:hamiltonian}. The functions $A(s)$ and $B(s)$, whose values for the D-Wave 2000Q machine at Los Alamos are given in Figure~\ref{fig:anneal_schedule}, control the anneal path. The parameter $s\in[0,1]$ is called \textit{anneal fraction}. At $s=1$ we have $A(s)=0$, and the final Hamiltonian describes a low-energy quantum system whose qubits'  values can be used to get a good solution of \eqref{eq:hamiltonian}.

The newest model of D-Wave annealers, 2000Q, gives more freedom to the user to tune some of the annealing control parameters. One of them is the \textit{anneal schedule} which, using up to $50$ points, specifies how the anneal fraction  evolves from the start to the end of the anneal \cite[Figure~2.1]{TechnicalDescriptionDwave}. Previously, this feature has been used for improving the accuracy of the annealer by inserting a pause is the anneal schedule \cite{Power_of_Pausing,PhysRevB.100.024302}.

The aim of this article is to investigate how the solution obtained by the D-Wave annealer changes during the anneal process. This process is unobservable, since D-Wave only allows users to read off the final qubit states at the end of each anneal. For this reason, we make use of a technique that allows us to approximate what a solution looks like at a given intermediate stage during the anneal. Using a custom anneal schedule, we follow the usual anneal curve up to the time at which we would like to obtain an intermediate solution. We then modify the anneal schedule by inserting a jump to the full anneal, thus \textit{freezing} the solution at the intermediate time point. This process, referred to as \textit{quenching}, and is a standard feature provided by the D-Wave 2000Q, see \cite[Section~2.5.2]{TechnicalDescriptionDwave}. Repeating this technique for various intermediate points allows us to \textit{slice} the anneal process.

We employ our technique to obtain various insights into the D-Wave 2000Q which, to the best of our knowledge, have not been reported previously in the literature. First, our approach allows us to visualize how the  solution evolves during the anneal process, both in terms of its energy and in terms of the total number of bit flips, that is, we measure how volatile the measured value of each individual qubit is during the anneal. This in turn allows us to determine the \textit{freeze-out point}, which is a point at which either the energy does not significantly improve anymore or, at an individual qubit level, the point at which the state/value of each individual qubit stays fixed. We repeat those experiments for a varying number of slices and annealing times.

Second, in order to better understand what happens during an anneal, we present a simple genetic optimization scheme \cite{MitchellGeneticAlgorithms} designed to find a QUBO that exhibits substantial improvements both in terms of energy decrease and number of bit flips during the anneal. We contrast the optimized QUBO with a random QUBO. Moreover, we study the characteristics and best parameter choices for our genetic algorithm.

Third, our technique allows us to track the freeze-out process of each individual qubit during the anneal. For this, we slice the anneal at various stages (for instance, using $1000$ slices for a $1000$ microsecond anneal), and observe how its measured value (zero or one) changes over the course of the anneal. The freeze-out point can then be defined as the approximate location at which the measured value of a qubit stays invariant until the end of the anneal.

The article is structured as follows. Section~\ref{sec:methods} presents our approach to slice the anneal process, that is the anneal schedule we employ in order to abort the anneal process at any intermediate time. We also present our genetic algorithm to find a QUBO that exhibits a substantial change during the anneal, both in terms of the total number of bit flips of all involved qubits, and the total energy decrease during the anneal. Using the optimized QUBO, Section~\ref{sec:simulations} will visualize how the solution evolves during the anneal, both in terms of number of bit flips and total energy change. We investigate how a random QUBO compares to our optimized QUBO, and how the number of slices influences the results. Importantly, the simulations demonstrate how our method can be used to visualize how individual qubits change during the anneal. A discussion of our methodology and results can be found in Section~\ref{sec:discussion}.

\section{Methodology}
\label{sec:methods}
This section describes the methodology underlying our work, consisting of two parts: First, Section~\ref{sec:slicing}  describes the technique we use in order to approximate the current state of the solution during the anneal process. To visualize results later in the simulations, we are interested in finding a QUBO (see Section~\ref{sec:intro}) that exhibits a pronounced evolution during the anneal process, that is (1) whose lowest energy result from a 1 microsecond D-Wave anneal is (significantly) greater than that of a longer (1000 microsecond) anneal and, (2) whose intermediate solutions involve a large number of bit flips during the anneal in order to arrive at the global minimum. We find such a QUBO with the help of a genetic algorithm presented in Section~\ref{sec:genetic_algo}.

\subsection{Slicing the anneal process}
\label{sec:slicing}

\begin{figure}
    \centering
    \vspace{-9mm}
    \includegraphics[width=0.8\textwidth]{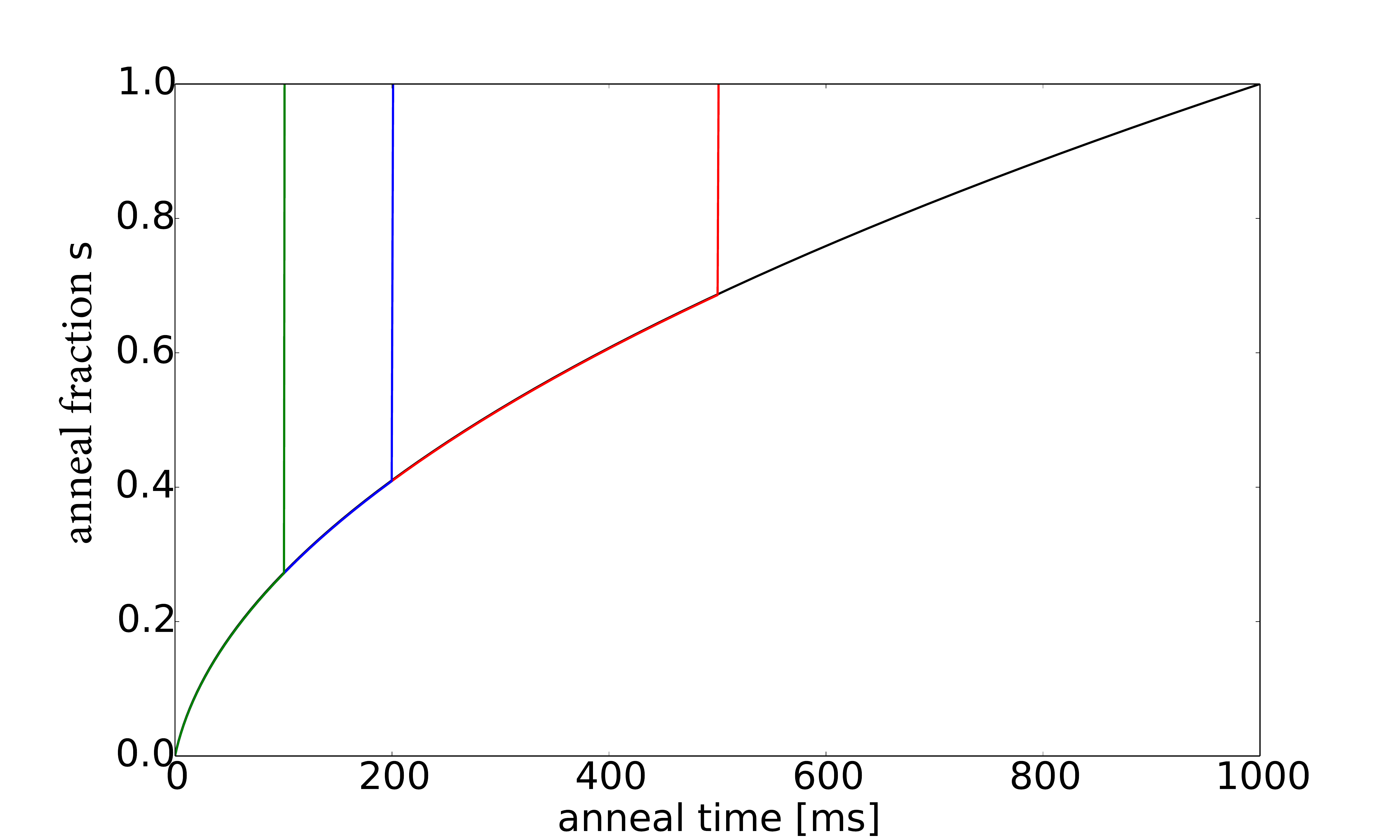}
    \caption{Anneal schedule with quenching (near-vertical jumps) at time points 100 (green), 200 (blue), and 500 (red) microseconds. Total anneal time of 1000 microseconds.}
    \label{fig:quenching}
\end{figure}

The goal of this work is to investigate how the anneal solution evolves during the anneal process of  D-Wave 2000Q. For this, we use a feature provided in the newest generation of D-Wave machines: it allows users to manually define an anneal schedule through the specification of up to $50$ points on the anneal curve.

Figure~\ref{fig:quenching} illustrates the anneal curves we employ. In order to slice the anneal process, we follow the standard anneal curve up to a time $t$ at which we would like to slice. Due to the hardware constraints of D-Wave 2000Q for specifying the anneal curve (precisely, D-Wave requires the maximal degree of the anneal curve to not exceed 45 degrees) as well as for consistency across all slices, we jump to the full anneal fraction of $1$ at time $t+1$.

Ideally, the jump from an intermediate anneal fraction to the end of the anneal has to be done as quickly as possible in order to "freeze" the solution at the anneal slice. Though the jump we employ is not perfectly vertical (our anneal curve jumps to $1$ at $t+1$, where $t$ is the slice we would like to observe), we expect this jump to not considerably change the solution, due in large part to the properties our genetic algorithm optimizes for. For consistency, we define the last anneal slice to be the full anneal.

Other anneal schedules are possible. For instance, one could pause the anneal at the current anneal fraction of slice $t$ (corresponding to e.g.\ roughly an anneal fraction of $0.4$ for slice $200$ in Figure~\ref{fig:quenching}), thus leaving it unchanged until the end of the anneal before jumping to $1$. We investigated such schedules but since results are qualitatively similar, we decided to only present our  simplest slicing technique, which exclusively employs quenching.

\subsection{A genetic algorithm to find a QUBO for better investigation of the anneal process}
\label{sec:genetic_algo}
It is known that different QUBOs pose different levels of difficulty to the D-Wave annealer when it comes to finding a high quality solution \cite{Chapuis2019}. In our experiments presented in Section~\ref{sec:simulations}, we aim to investigate how the anneal solution changes during the anneal process, and therefore seek to solve a non-trivial QUBO instance that exhibits a distinctive evolution measured both in terms of energy decrease and bit flips during the anneal process.

To find such a QUBO, we employ the genetic algorithm that works as follows. The algorithm starts by initializing a population of $N \in \N$ QUBOs which are collected in a set $S$. Those QUBOs are randomly generated as follows: each QUBO fills the entire chimera graph, its linear weights are independently sampled from $(-2,2)$ and its couplers are independently sampled from $(-1,1)$, the ranges of weights allowed by the hardware.

Then the algorithm proceeds by evaluating the fitness of the current population of QUBOs. To this end, for each $Q \in S$, we find the minimum energy solution across $1000$ D-Wave anneals for a $1$ microsecond and a $1000$ microsecond anneal. We then compute the energy difference $\Delta$ between both. Moreover, we find the \textit{Hamming distance} (the number of bit positions with different values) $d$ between the minimum energy solutions for a full $1000$ microsecond anneal and a $1$ microsecond slice (using the slicing technique of Section~\ref{sec:slicing}). We set the fitness $f_Q$ for each $Q \in S$ to $\Delta \cdot (d/n \cdot 100)$, where $n$ is the number of qubits in $Q$. Hence, the fitness is the energy decrease between the $1$ and $1000$ microsecond anneals multiplied by the percentage of bit flips between the $1$ microsecond slice and the full anneal. This ensures that the genetic algorithm ~\ref{sec:genetic_algo} will optimize for QUBOs having the property that between a $1$ microsecond and a full $1000$ microsecond anneal, their minimum energy solution evolves considerably both in terms of energy and actual bit assignment.

Next, the proportion of $p_\text{cross}$ fittest individuals are selected from the population for cross-over and mutation. Those QUBOs are stored in the set $S_0$. Then we restore the original size $N$ of the population by crossing the fittest individuals from set $S_0$: we randomly choose two QUBOs $Q_1$ and $Q_2$ from $S_0$ and generate a new QUBO by selecting each individual weight and quadratic coupler independently from either $Q_1$ or $Q_2$ with probability $0.5$. We store the new QUBO in $S_1$ and repeat this step $N$ times.

Finally, a mutation step is applied to the new population. For each QUBO from the new population, $Q \in S_1$, we overwrite any of the coefficients of $Q$ with probability $p_\text{mut}$. As in the initialization step, a linear weight is sampled randomly from $(-2,2)$ and a quadratic coupler is sampled from $(-1,1)$. The unmutated and mutated QUBOs are stored in a new set $S_2$. After setting $S \leftarrow S_2$, the genetic algorithm is restarted with the newest population.

The entire process is repeated over $R$ iterations. After the last iteration, we return the fittest $Q \in S$ as the result of our algorithm, where $S$ is the newest population generation and fitness is calculated using function $f_Q$.

The dependence of the genetic algorithm described in ~\ref{sec:genetic_algo} on its parameters is evaluated in Section~\ref{sec:GA_parameter_choice}, where we also suggest default parameter choices.

\section{Experiments}
\label{sec:simulations}
This section presents our experimental setup and results. We start with an assessment of the tuning parameters of our genetic algorithm (Section~\ref{sec:GA_parameter_choice}). In Section~\ref{sec:random_vs_optimized}, we present first results on the slicing technique of Section~\ref{sec:slicing}, demonstrating that the optimized QUBO indeed yields more pronounced changes during the anneal. The remaining sections give insights into how the energy (Section~\ref{sec:evolution_energy}) and the Hamming distance (Section~\ref{sec:evolution_hamming}) of the best solution evolves during annealing. Section~\ref{sec:freezeout_chimera} shows freeze-out points for all individual qubits during annealing.

Apart from Section~\ref{sec:random_vs_optimized}, which also displays results for a random QUBO, all figures in the remainder of the simulations were computed using an optimized QUBO obtained with the genetic algorithm described in \ref{sec:genetic_algo}.

The results of the following subsections are reported with error bars, with the exception of Figure~\ref{fig:1000_rand_opt}. We compute those results as follows: We run $1000$ anneals on D-Wave 2000Q and record the $10$ best (minimal) energies found among those $1000$ anneals. We repeat this step $10$ times, thus resulting in $100$ measurements. We then plot the mean of those measurements together with one standard deviation error bar.

\subsection{Parameter choices for the genetic algorithm}
\label{sec:GA_parameter_choice}
We start by assessing the dependence of the genetic algorithm described in ~\ref{sec:genetic_algo} on its tuning parameters. For this we first define a set of default parameters: population size $N=50$, crossover proportion $p_\text{cross}=0.25$, and mutation rate $p_\text{mut}=0.01$.

\begin{figure*}
    \centering
    \includegraphics[width=0.49\textwidth]{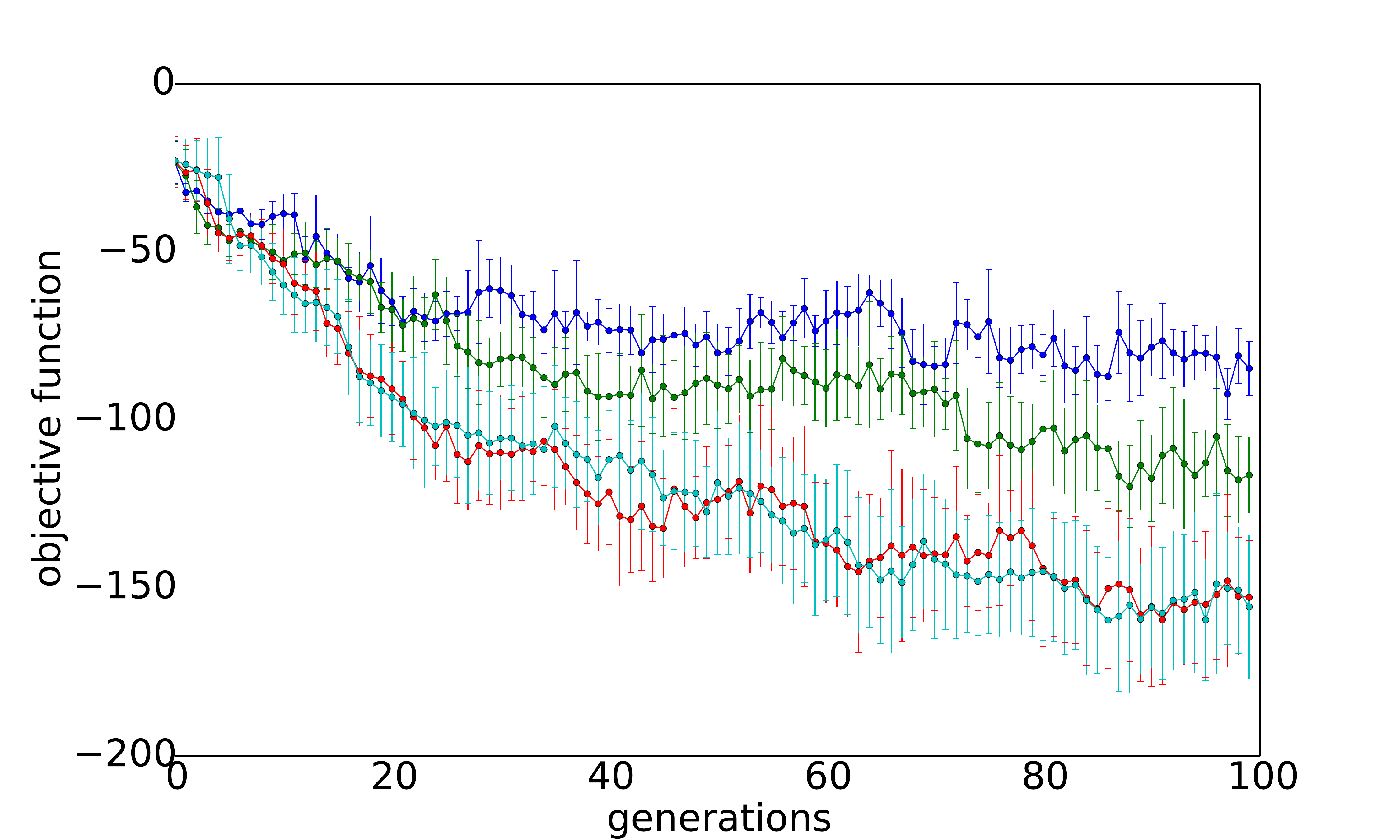}\hfill
    \includegraphics[width=0.49\textwidth]{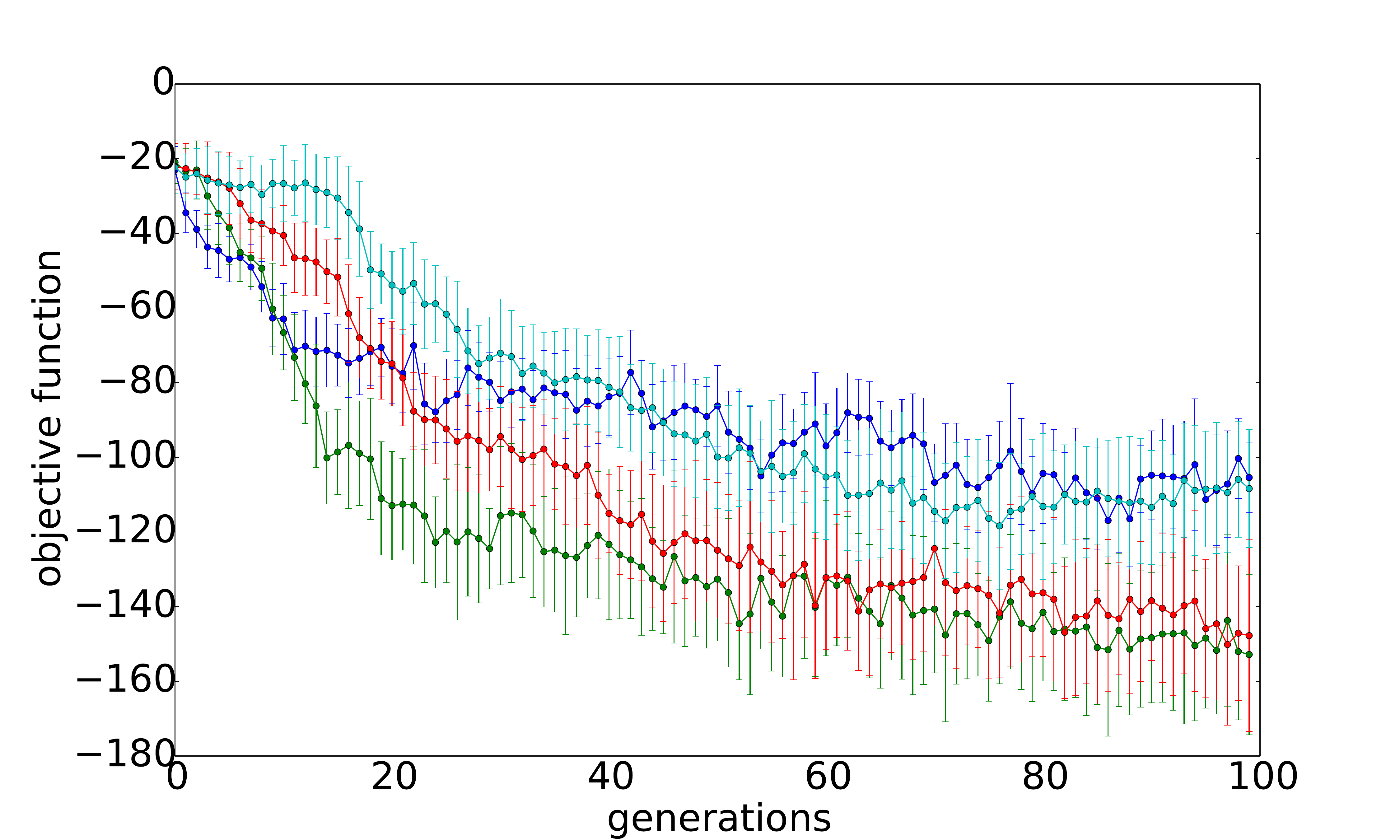}\\
    \includegraphics[width=0.49\textwidth]{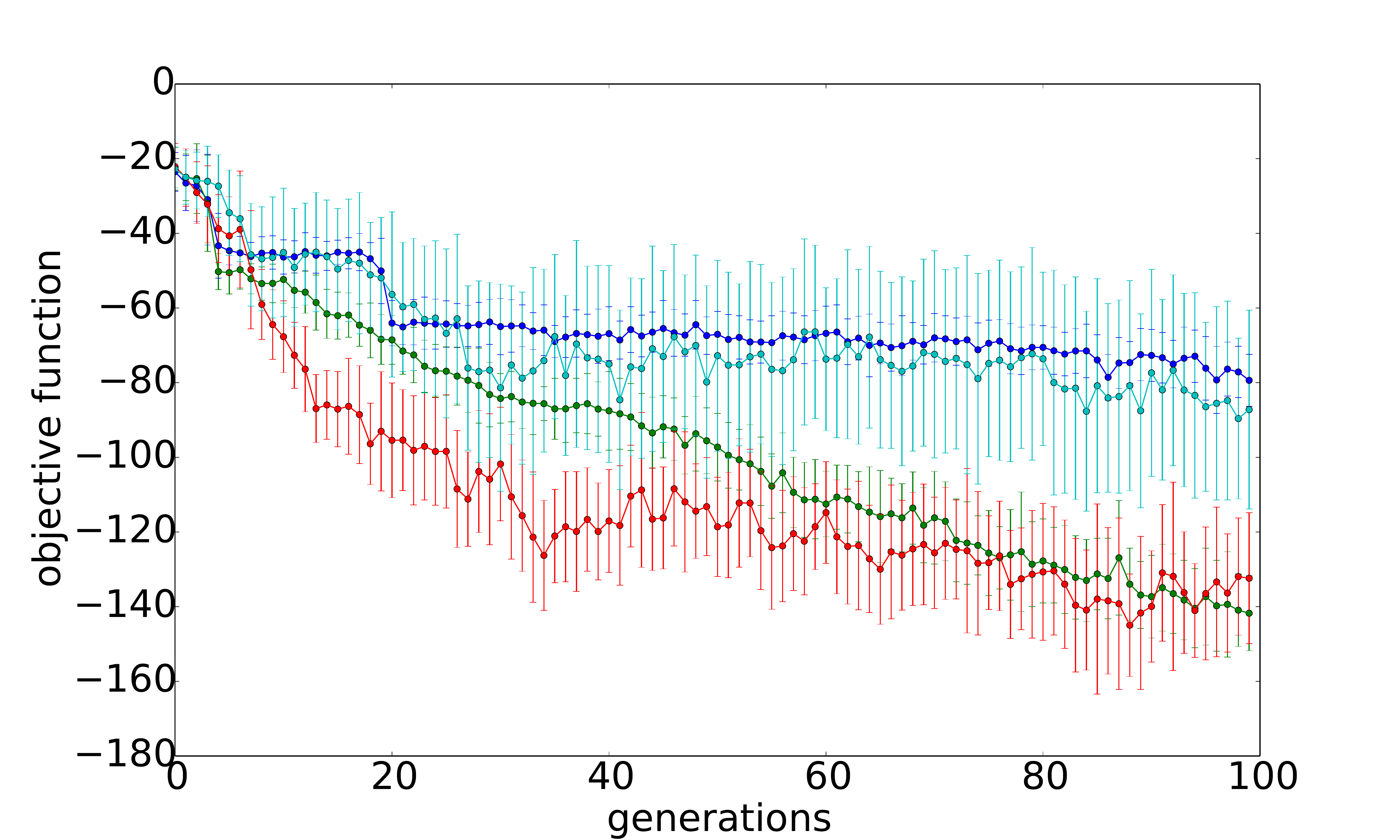}\hfill
    \includegraphics[width=0.49\textwidth]{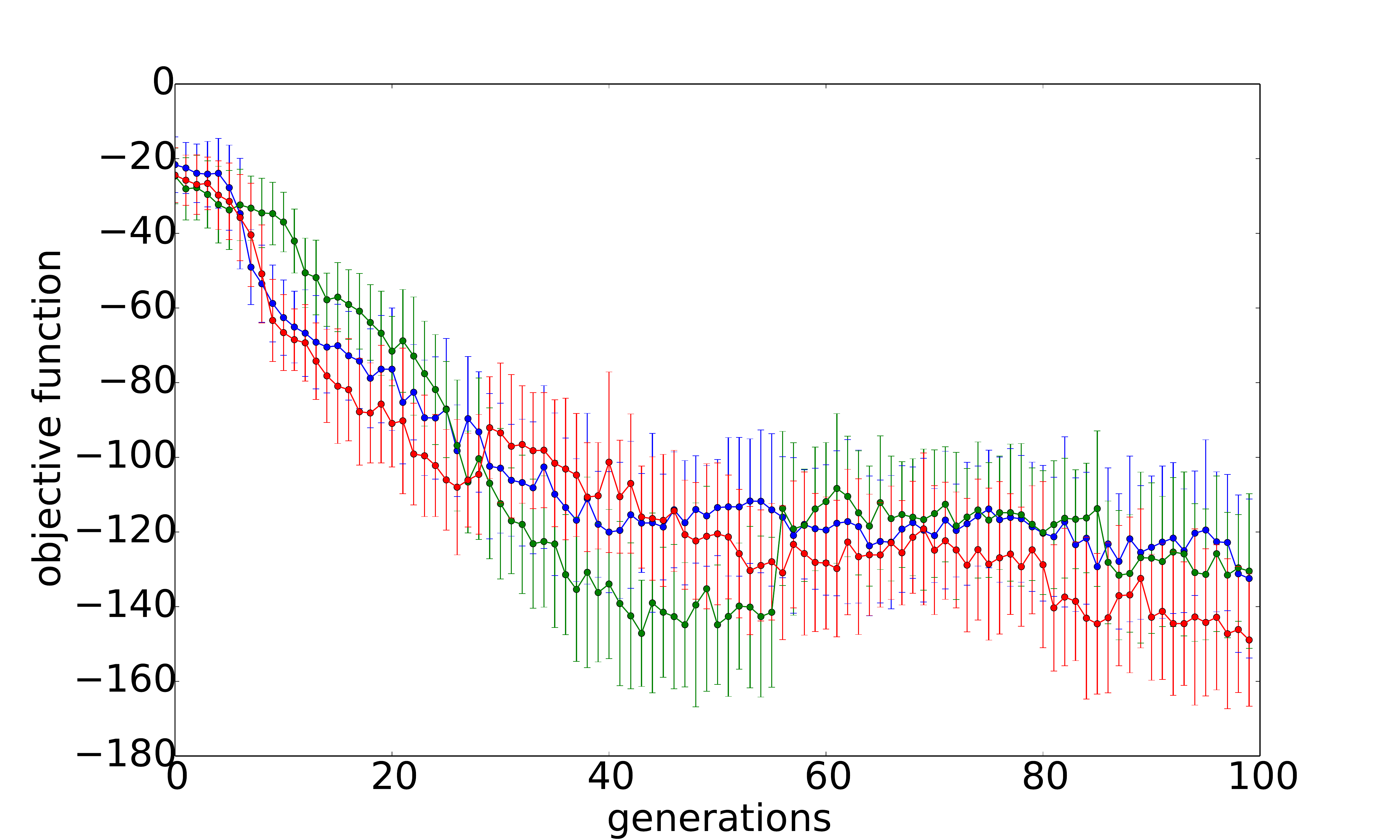}
    \caption{Dependence of the genetic algorithm on its tuning parameters while keeping the other parameters at their default values. Population size $N \in \{10,25,50,75\}$ (top left), crossover proportion $p_\text{cross} \in \{0.1,0.25,0.5,0.75\}$ (top right), mutation rate $p_\text{mut} \in \{0.0001,0.001,0.01,0.1\}$ (bottom left). The colors blue, green, red, cyan always indicate the four parameter choices from smallest to largest value. Bottom right shows three runs of the genetic algorithm with the optimized parameters.}
    \label{fig:genetic_algo}
\end{figure*}

While keeping two of the default parameters fixed, we vary the third parameter in Figure~\ref{fig:genetic_algo}. The figure shows that increasing the population size above $N=50$ does not considerably change the behavior of the algorithm. Similarly, using a crossover rate of around $p_\text{cross}=0.25$ seems to be advantageous, and likewise a too low or too high mutation rate is disadvantageous, thus leading us to the choice $p_\text{mut}=0.01$. The default parameters will be used in the remainder of the simulations. Figure~\ref{fig:genetic_algo} (bottom right) shows three runs of the genetic algorithm with default parameters, demonstrating that around $50$ iterations are needed to find a QUBO with the desired properties.

\subsection{Evolution of solutions for random vs.\ optimized QUBO}
\label{sec:random_vs_optimized}
\begin{figure}
    \centering
    \includegraphics[width=0.7\textwidth]{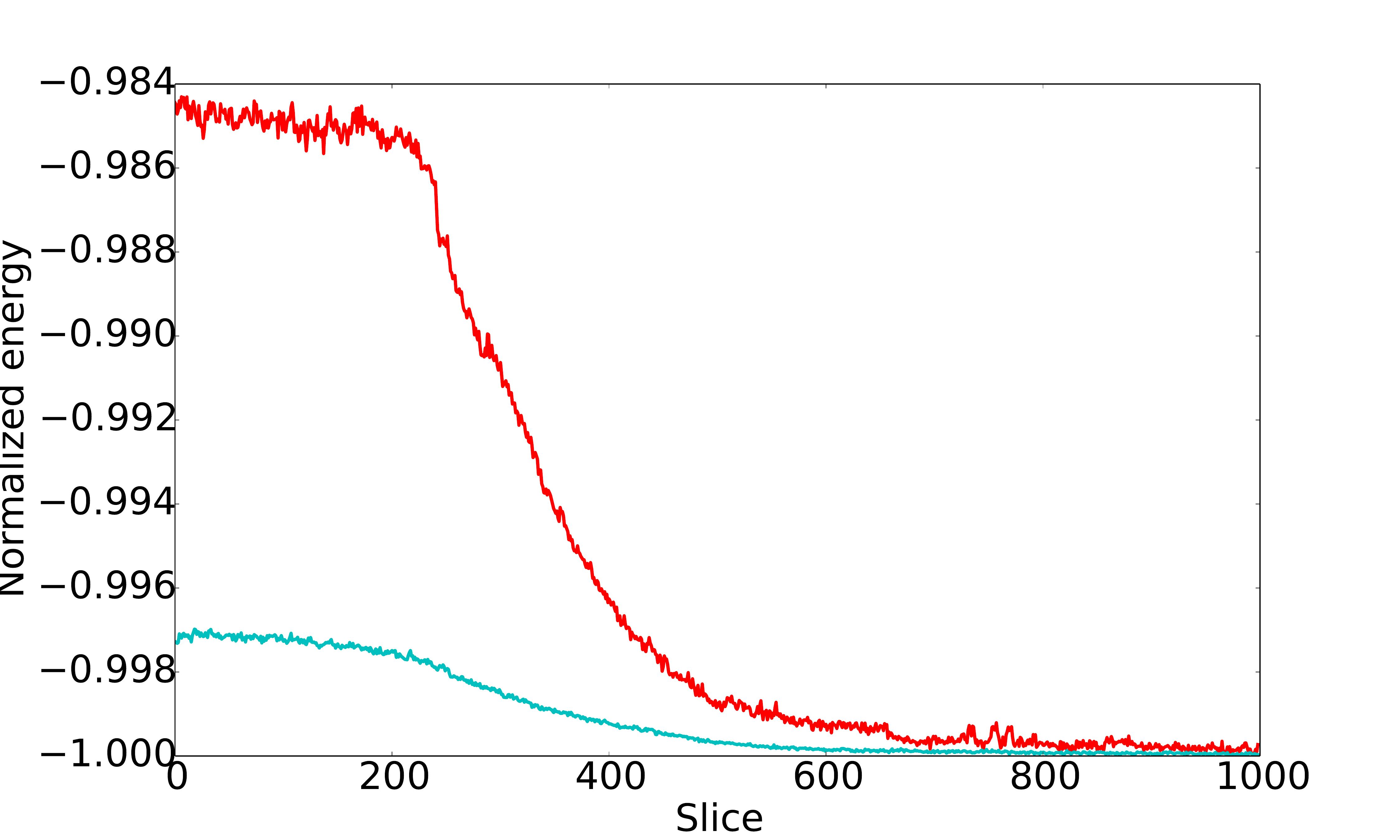}
    \caption{Averages of the minimum 1 percent energies found in 10 runs of 1000 anneals. Random QUBO (cyan) and optimized QUBO (red). Both datasets are normalized by dividing each value by their respective minimum value.}
    \label{fig:1000_rand_opt}
\end{figure}

We verify that using a QUBO computed with the genetic algorithm~\ref{sec:genetic_algo} is indeed advantageous for investigating the evolution of the solution during the anneal. Figure~\ref{fig:1000_rand_opt} shows, for $1000$ slices and $10$ averages for each slice, the progression of the average of the $1$ percent minimum energies found.

Figure~\ref{fig:1000_rand_opt} shows that, the optimized QUBO requires a longer annealing time in order to gain a significant benefit in the form of a large decrease in the best energies found. Of significance is that that the total change in energy is significantly more for the optimized QUBO. We suggest defining the freeze-out point at roughly slice $600$ where both curves stabilize horizontally.

\subsection{Energy evolution during the anneal process}
\label{sec:evolution_energy}
\begin{figure*}
    \centering
    \includegraphics[width=0.49\textwidth]{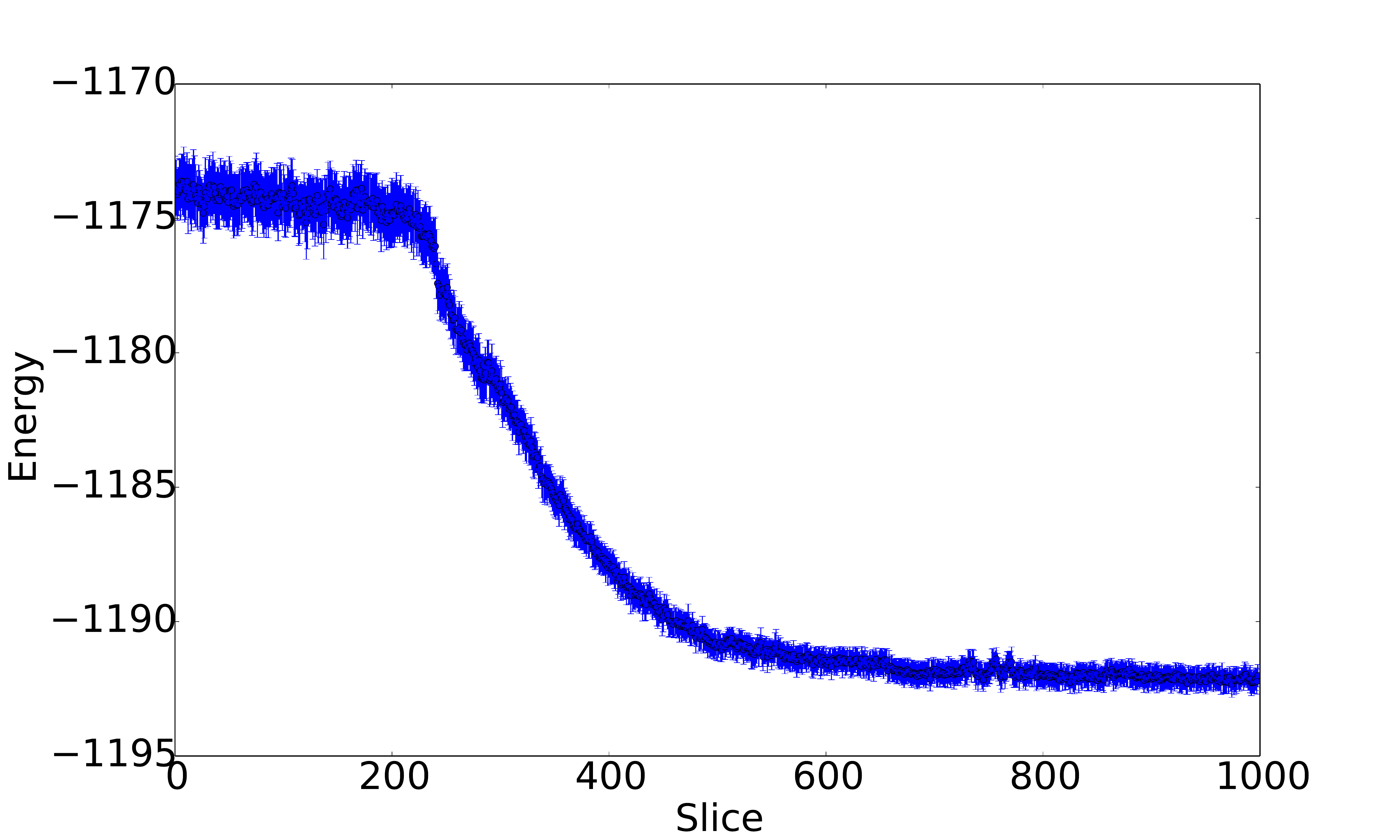}\hfill
    \includegraphics[width=0.49\textwidth]{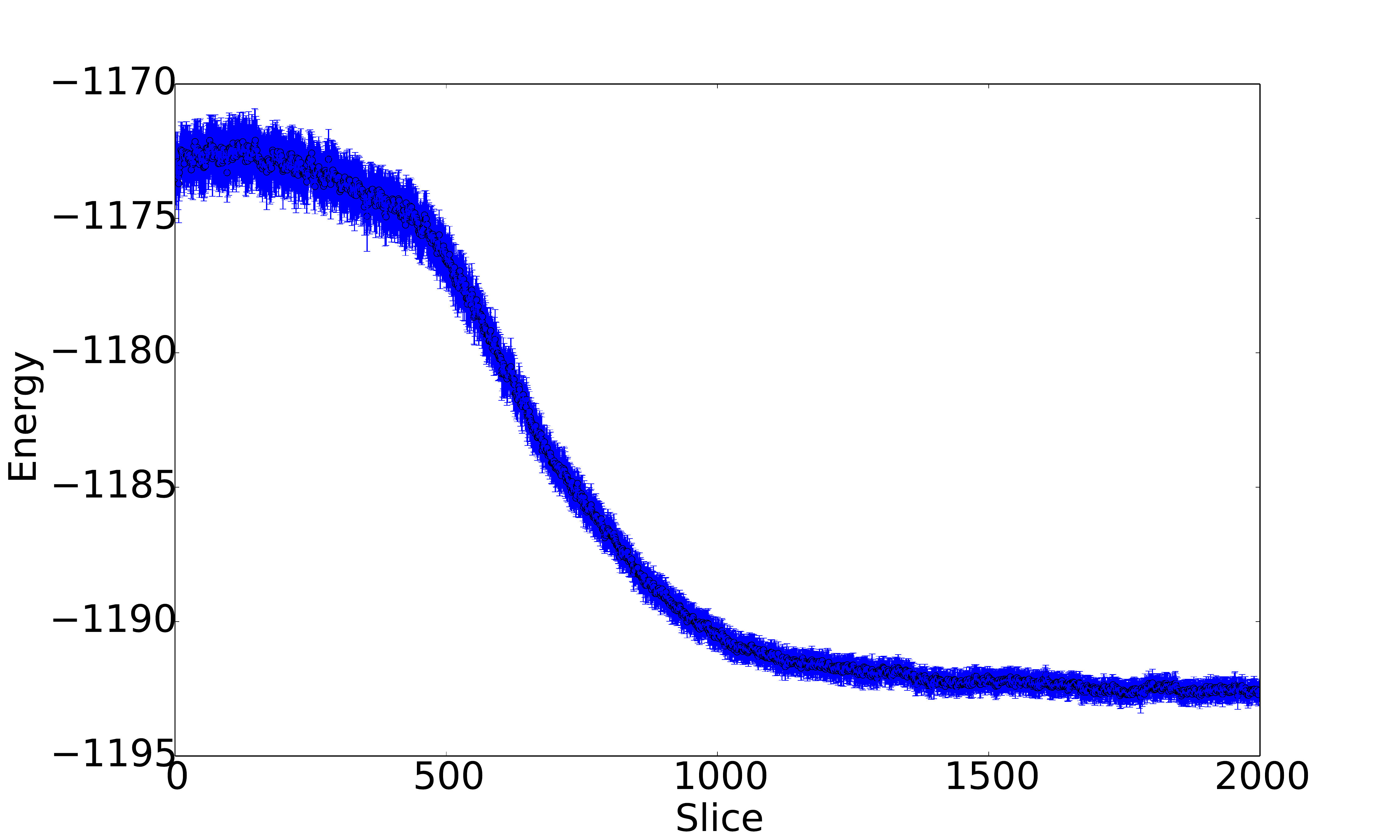}
    \caption{Evolution of minimum energy solution on the D-Wave 2000Q for 1000 slices (left) and 2000 slices (right).}
    \label{fig:energies}
\end{figure*}

We now look at the evolution of energies for the optimized QUBO. Figure~\ref{fig:energies} displays results for the optimized QUBO when dissecting the anneal process using $1000$ slices of a $1000$ microsecond anneal (left) as well as $2000$ slices of a $2000$ microsecond anneal (right). We observe that, initially, the energies only decrease slowly up to around a quarter of the anneal. At that point, a continuous reduction in energies sets in, during which the anneal solution becomes better and better with every slice. Moreover, we also observe that the error on the energy readings becomes smaller at that point, meaning that the energy distribution of the anneals returned by D-Wave is considerably  narrower.

\subsection{Evolution of the Hamming distance between adjacent slices}
\label{sec:evolution_hamming}
Similarly to Figure~\ref{fig:energies}, we repeat the experiment and record the Hamming distance between the binary solution vectors (indicating the final measured value of each qubit) of adjacent slices.

\begin{figure*}
    \centering
    \includegraphics[width=0.49\textwidth]{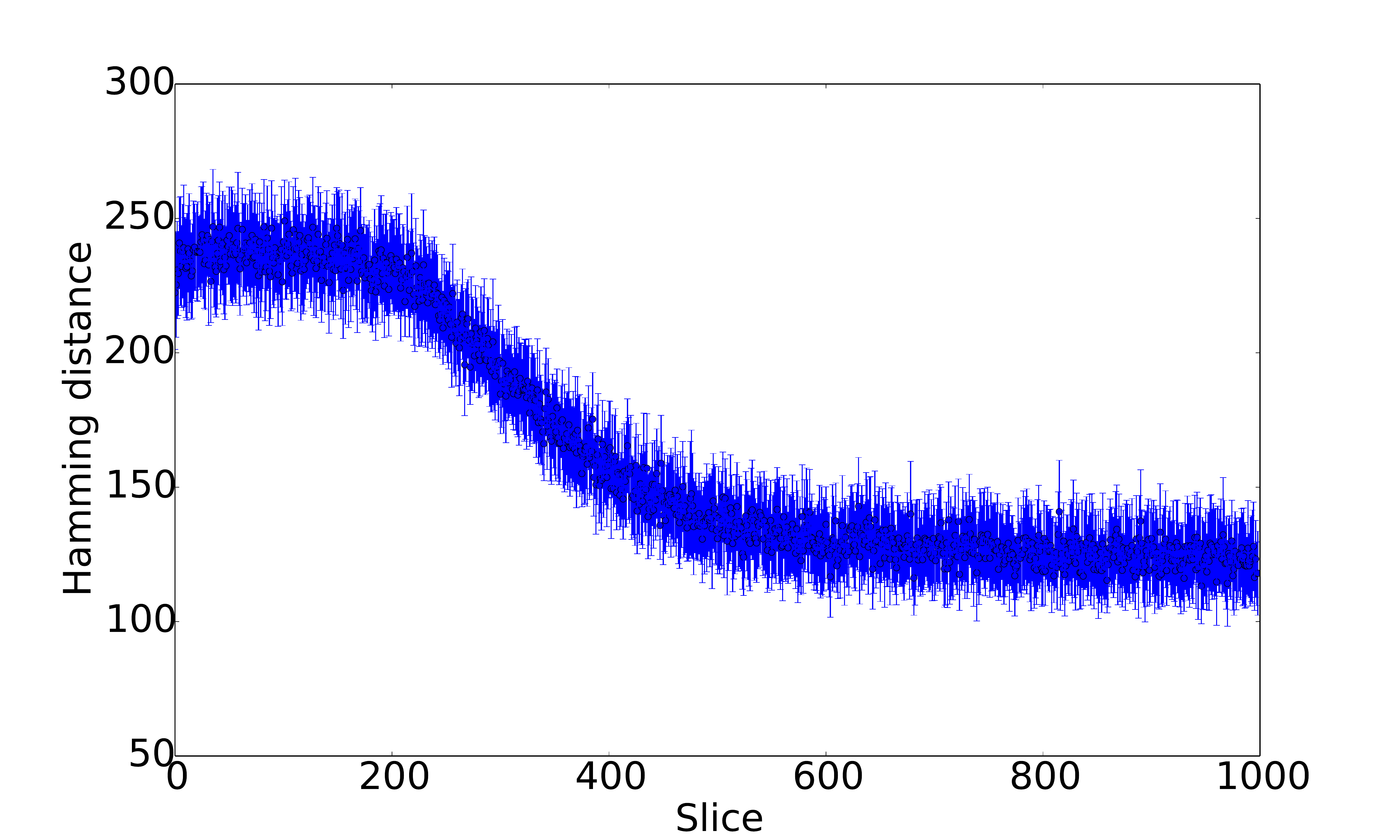}\hfill
    \includegraphics[width=0.49\textwidth]{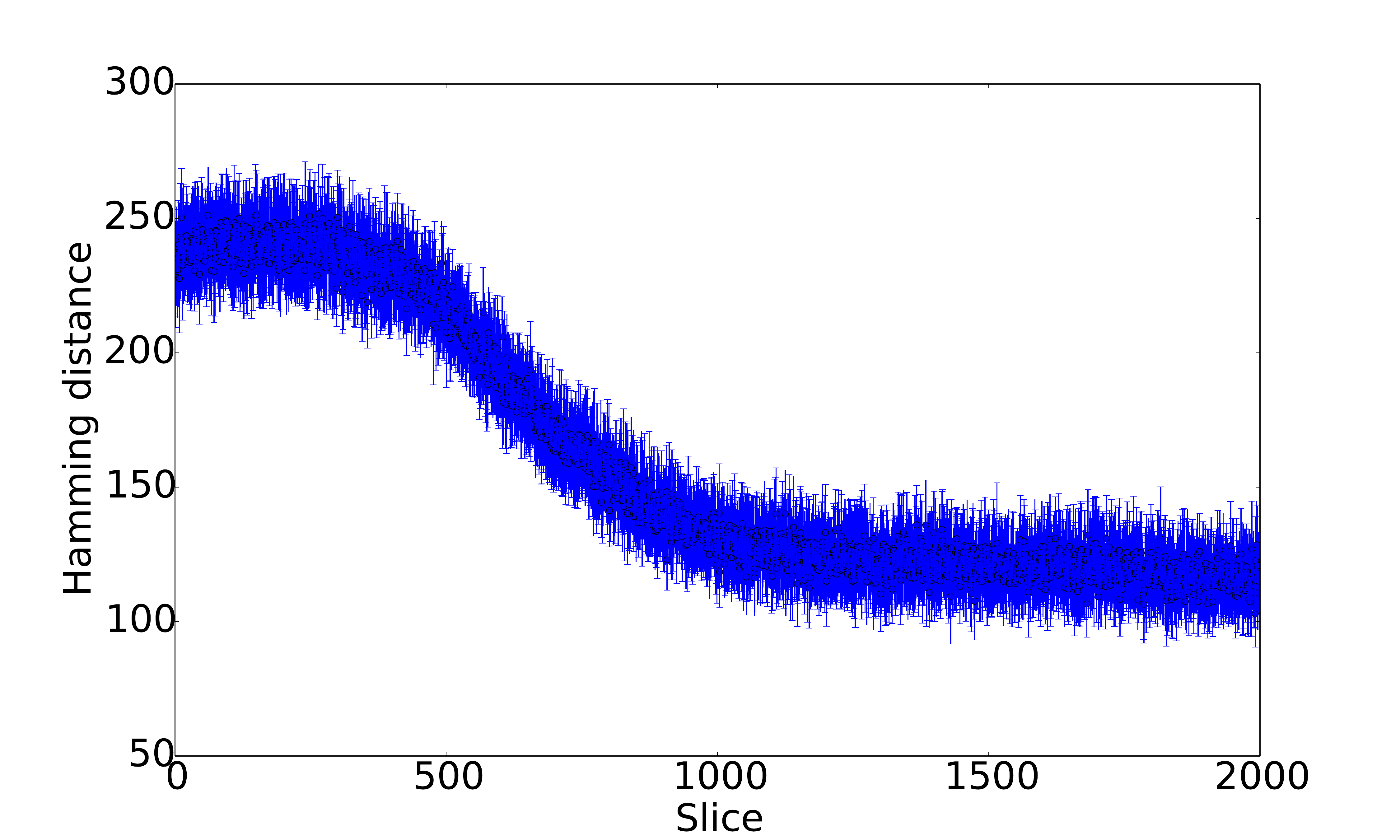}
    \caption{Evolution of the Hamming distance for the minimum energy solution between adjacent slices. Left: 1000 slices. Right: 2000 slices.}
    \label{fig:hamming_distance}
\end{figure*}

Figure~\ref{fig:hamming_distance} shows the evolution of the Hamming distance for the minimum energy solution between adjacent slices, both for $1000$ slices of $1000$ microsecond annealing time (left) and $2000$ slices of $2000$ microsecond annealing time (right). As expected, the shapes of the curves are similar in the two cases. However, several interesting observations can be made: First, while the energy of the solutions initially stays almost constant and then shows a pronounced decline during the anneal (Figure~\ref{fig:energies}), this decrease is much less pronounced when looking at the number of bit changes of the solution vectors. Second, the freeze-out points (defined as the time at which the energy or Hamming distance plateaus) determined from the evolutions of either the energy (Figure~\ref{fig:energies}) or Hamming distance roughly coincide.

\subsection{Determining the freeze-out point for each individual qubit}
\label{sec:freezeout_chimera}
Finally, we can use our methodology from Section~\ref{sec:slicing} to track the freeze-out points of individual qubits of  D-Wave. For this end, we again employ the optimized QUBO obtained with the genetic algorithm (see Section ~\ref{sec:genetic_algo}), and read out the value of each qubit at each of 1000 slices during the anneal process. We visualize the rate with which each bit flips in adjacent slices with different colors: blue indicates no bit flips (the qubit kept the same measured value from the start to the end of the anneal), red indicates a high rate of bit flips, and weaker colors show intermediate rates.

\begin{figure}
    \centering
    \includegraphics[width=\textwidth]{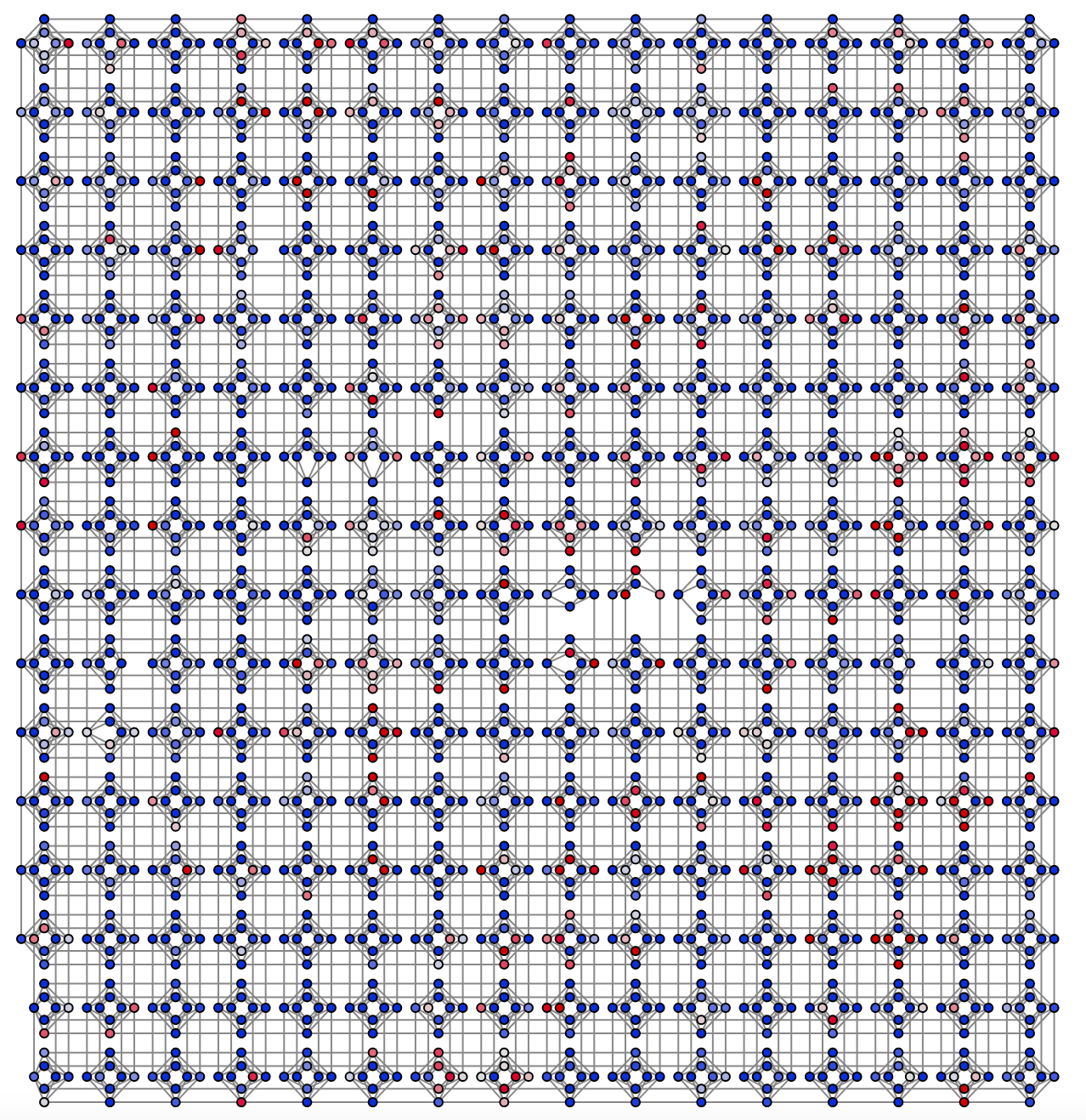}
    \caption{Rate of bit flips of individual qubits on the D-Wave chimera graph. Colors indicate rate of bit flips from zero (blue) to high rate (red).}
    \label{fig:freezeout_chimera}
\end{figure}

Figure~\ref{fig:freezeout_chimera} shows results for the LANL D-Wave 2000Q chimera graph. We observe that although the QUBO employed for annealing used up all linear weights and quadratic couplers on the chip, the majority of qubits remained in their initial (i.e. first slice) measured value (either zero or one) during the anneal. It also seems as if qubits with either low or high rates of bit flips somehow cluster together. This is expected as it is known that there is leakage on the chip, meaning that adjacent qubits influence each other: thus it is sensible that a qubit which changes at a high rate during the anneal also induces bit flips in adjacent qubits.

\section{Discussion}
\label{sec:discussion}
This article is a first attempt to explore how the solution obtained by the D-Wave 2000Q quantum annealer evolves during the anneal process. To the best of our knowledge, such work has not been presented previously in the literature.

In order to estimate the states of individual qubits at any point during the anneal, we develop a novel method we refer to as slicing, which uses the 2000Q quenching control feature. Using this technique, we dissect the anneal process and monitor both how the energy of the solution evolves, and how the state of each qubit changes from slice to slice.

We summarize our findings as follows:
\begin{enumerate}
    \item We demonstrate that using an optimized QUBO computed with a genetic algorithm results in a much more pronounced evolution during the anneal than using a random QUBO.
    \item Using the optimized QUBO, we observe that, initially, the energy does not considerably decrease during the anneal. At roughly a quarter of the anneal, a pronounced decrease sets in, which correlates with a reduction in the number of bit flips. At around the midpoint of the anneal, the energy stabilizes at around the energy value that D-Wave returns as a solution at the end of the anneal. During that phase, the number of bit flips likewise stabilizes at a constant level (meaning that the solution still changes from slice to slice) which could reflect the noise in the machine.
    \item We show that many qubits actually keep their initial measured value during the anneal, and that qubits with a high rate of bit flips during the anneal seem to cluster together.
\end{enumerate}

This work serves as a first step in the development of methods that allows getting insights into the (unobservable) anneal process. Indeed, further avenues of research are possible:
\begin{enumerate}
    \item Investigating the potential utility of measuring individual qubit freezeout points.
    \item Slicing anneals with shorter total annealing times and smaller annealing time increments compared to the results we present. 
    \item Using this slicing technique we present in order to look at the evolution of specific optimization problems, such as the Maximum Clique problem.
    \item Analyzing slices of chained (minor embedded) problems in order to determine freeze out of chains relative to all active variables on the chip.
\end{enumerate}

\bibliographystyle{apalike}

\end{document}